\documentstyle[prl,twocolumn,aps,floats,epsfig]{revtex}

\begin{document}

\draft
\twocolumn[\hsize\textwidth\columnwidth\hsize\csname
@twocolumnfalse\endcsname

\title{Principles of superconductivity} 

\author{A. Mourachkine} 

\address{Nanoscience Centre, 
University of Cambridge, 11 J. J. Thomson Avenue, 
Cambridge CB3 0FF, UK} 

\maketitle

\begin{abstract} 

The purpose of this chapter is to discuss the main principles of 
superconductivity as a phenomenon, valid for every superconductor 
independently of its characteristic properties and material. The underlying 
mechanisms of superconductivity can be different for various 
materials, but certain principles must be satisfied. The chapter 
introduces four principles of superconductivity. (The chapter is slightly 
modified from the original one in order to be self-contained.)

\end{abstract}

\pacs{{\bf \footnotesize (Chapter 4 in a book {\it Room-Temperature
Superconductivity}  (Cambridge International Science Publishing, Cambridge, 2004))}} 
]

\vspace*{2mm} 

The issue of room-temperature 
superconductivity is the main topic of this book. Even if this subject 
was raised for the first time before the development of the BCS 
theory and later by Little in 1964  [1], from the 
standpoint of practical realization, this issue is still a new, 
``untouched territory.'' To go 
there, we need to know Nature's basic rules for arrangement of matter 
over there. Otherwise, this journey will face a {\em fiasco}. To have 
the microscopic BCS theory in a bag is very useful, but not 
enough. It is clear to everyone by now that a room-temperature superconductor 
can not be of the BCS type. Therefore, we need to know more general 
rules, principles of superconductivity that incorporate also the 
BCS-type superconductivity as a 
particular case. 

The purpose of this chapter is to discuss the main principles of 
superconductivity as a phenomenon, valid for every superconductor 
independently of its characteristic properties and material. The underlying 
mechanisms of superconductivity can be different for various 
materials, but certain principles must be satisfied. One should however 
realize that the principles of superconductivity are not limited to those 
discussed in this chapter: it is possible that there are others which we 
do not know yet about. 

The first three principles of superconductivity were introduced in [2]. 

\section{First principle of superconductivity} 

The microscopic theory of superconductivity for {\em conventional} 
superconductors, the BCS 
theory, is based on Leon Cooper's work published 
in 1956. This paper was the first major breakthrough for understanding 
the phenomenon of superconductivity on a {\em microscopic} scale. Cooper 
showed that electrons in a solid would always form pairs if an attractive 
potential was present. It did not matter if this potential was very weak. 
It is interesting that, during his calculations, Cooper was not looking for 
pairs---they just ``dropped out'' of the mathematics. Later it became 
clear that the interaction of electrons \\ \vspace{2mm}

with the lattice allowed them to 
attract each other despite their mutual Coulomb repulsion. These electron 
pairs are now known as Cooper pairs. 

$\underline{{\rm An \ important \ note}}$: 
in this chapter, we shall 
use the term ``a Cooper pair'' more generally than its initial meaning. In the 
framework of the BCS theory, the Cooper pairs are formed in 
momentum space, not in real space. Further, we shall consider the case of 
electron pairing in real space. For simplicity, we shall sometimes call 
electron pairs formed in real space also as Cooper pairs. 

In solids, superconductivity as a quantum state cannot occur without the 
presence of bosons. 
Fermions are not suitable for forming a quantum state 
since they have spin and, therefore, they obey the Pauli exclusion principle 
according to which two identical fermions cannot occupy the same quantum 
state. Electrons are fermions with a spin of 1/2, while Cooper 
pairs are already composite bosons since the value of their total spin is 
either 0 or 1. Therefore, the electron pairing is an inseparable part of the 
phenomenon of superconductivity and, in any material, superconductivity 
cannot occur without electron pairing. 

In some unconventional superconductors, the charge carriers are not 
electrons but holes with a charge of $+|e|$ and spin of 1/2. The 
reasoning used above for electrons is valid for holes as well. Thus, in the 
general case, it is better to use the term ``quasiparticles'' which also 
reflects the fact that the electrons and holes are in a medium. 

The first principle of superconductivity: 

\vspace{0.5cm} 
\noindent
Principle 1: 
{\bf Superconductivity requires \\ \hspace*{19mm} quasiparticle pairing} 
\vspace{0.5cm}

In paying tribute to Cooper, the first principle of superconductivity can be 
called the {\em Cooper principle}. 

In the framework of the BCS theory, the quasiparticle (electron) 
pairing occurs in momentum space, not in real space. Indeed in the next 
section, we shall see that the electron pairing in conventional 
superconductors cannot occur in real space because the onset of 
long-range phase coherence in classical superconductors occurs due 
to the overlap of Cooper-pair wavefunctions, as shown in Fig. 1. 
As a consequence, the order parameter and the Cooper-pair wavefunctions in 
conventional superconductors are the same: the order parameter is a 
``magnified'' version of the Cooper-pair wavefunctions. However, in 
unconventional superconductors, the electron pairing is not restricted 
by the momentum space because the order parameter in 
unconventional superconductors has nothing to do with the Cooper-pair 
wavefunctions. {\em Generally speaking}, the electron pairing in 
\begin{figure}[t]
\epsfxsize=0.95\columnwidth
\centerline{\epsffile{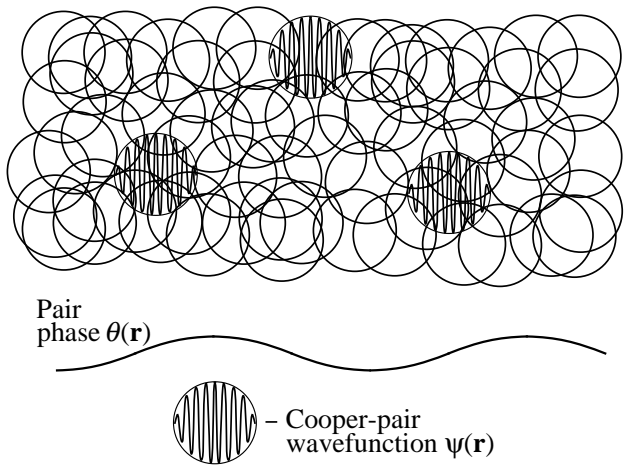}}
\vspace{5mm}
\caption{In conventional superconductors, 
the superconducting ground state is composed by a very large 
number of overlapping Cooper-pair wavefunctions, $\psi ({\bf r})$. 
To avoid confusion, only three Cooper-pair wavefunctions are shown in 
the sketch; the other are depicted by open circles. The phases of the 
wavefunctions are locked together since this minimizes the free energy. 
The Cooper-pair phase $\Theta ({\bf r})$, illustrated in the sketch, is 
also the phase of the order parameter $\Psi({\bf r})$.}
\end{figure}
unconventional superconductors may take place not only in momentum  
space but also in real space. We shall discuss such a possibility in the 
following section. 

The electron pairing in momentum space can be considered as a 
{\em collective} phenomenon, while that in real space as 
{\em individual}. We already know that the density of free (conduction) 
electrons in conventional superconductors is relatively high 
($\sim$ 5 $\times$ 10$^{22}$ cm$^{-3}$); however, only a small fraction 
of them participate in electron pairing ($\sim$ 0.01\%). In 
unconventional superconductors it is just the other way round: the 
electron density is low ($\sim$ $5 \times10^{21}$ cm$^{-3}$) but a 
relatively large part of them participate in the electron pairing 
($\sim$ 10\%). Independently of the space where they 
are paired---momentum or real---two electrons can form a bound state 
{\bf only if} {\em the net force acting between them is attractive}. 

\section{Second principle of superconductivity} 

After the development of the BCS theory in 1957, the issue of 
long-range phase coherence in superconductors was not discussed widely 
in the literature because, in conventional superconductors, the 
pairing and the onset of phase coherence take place simultaneously 
at $T_c$. The onset of phase coherence in conventional superconductors 
occurs due to the overlap of Cooper-pair wavefunctions, as shown in Fig. 1. 
Only after 1986 when high-$T_c$ superconductors were discovered, the 
question of electron pairing above $T_c$ appeared. So, it was then realized 
that it is necessary to consider the two processes---the electron pairing 
and the onset of phase coherence---separately and independently of one 
another [3]. 

In many unconventional superconductors, quasiparticles become paired 
above $T_c$ and start forming the superconducting condensate only at 
$T_c$. Superconductivity requires both the electron pairing and the 
Cooper-pair condensation. Thus, the second principle of superconductivity 
deals with the Cooper-pair condensation taking place at $T_c$. This 
process is also known as the onset of long-range phase coherence.

\vspace{0.5cm} 
\noindent
Principle 2: {\bf The transitioninto the superconduct- \\ \hspace*{19mm}ing 
state is the Bose-Einstein-like \\ \hspace*{19mm}condensation and occurs in
\\ \hspace*{19mm}momentum space} 
\vspace{0.5cm}

Let us first start with one main difference between fermions 
and bosons.  
Figure 2 schematically shows an ensemble of fermions and an ensemble 
of bosons at $T \gg$ 0 and $T$ = 0. In Fig. 2 one can see that, at high 
temperatures, both types of particles behave in a similar manner by 
distributing themselves in their energy levels somewhat haphazardly but 
with more of them toward lower energies. At absolute zero, the two types 
of particles rearrange themselves in their lowest energy configuration. 
Fermions obey the Pauli exclusion principle. Therefore, at absolute zero, 
each level from the bottom up to the Fermi energy $E_F$ 
is occupied by two 
electrons, one with spin up and the other with spin down, as shown in 
Fig. 2. At absolute zero, all energy levels above the Fermi level are empty. 
In contrast to this, bosons do not conform to the exclusion principle, 
therefore, at 
absolute zero, they all consolidate in their lowest energy state, as shown 
in Fig. 2. Since all the bosons are in the same quantum state, they form a 
quantum condensate (which is similar to a superconducting condensate). 
In practice, however, absolute zero is not accessible. 
\begin{figure}[t]
\epsfxsize=0.95\columnwidth
\centerline{\epsffile{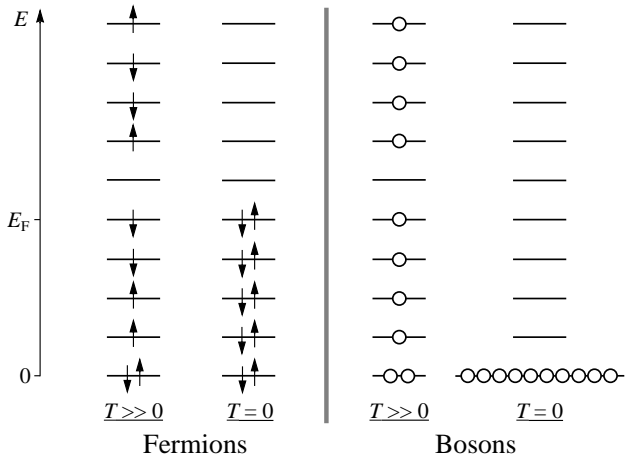}}
\vspace{5mm}
\caption{Sketch of the occupation of energy levels for fermions and 
bosons at high temperatures and absolute zero. Arrows indicate the 
spin direction of the fermions. For simplicity, the spin of the bosons is 
chosen to be zero. $E_F$ is the Fermi level for the fermions.}
\end{figure} 

We are now ready to discuss the so-called 
{\em Bose-Einstein condensation}. 
In the 1920s, Einstein predicted that if an ideal gas of identical atoms, i.e. 
bosons, at thermal equilibrium is trapped in a box, at sufficiently 
low temperatures the particles can in principle accumulate in the lowest energy 
level (see Fig. 2). This may take place only if the quantum wave packets of 
the particles overlap. In other words, the wavelengths of the matter waves 
associated with the 
particles---the {\em Broglie waves}---become similar 
in size to the mean particle distances in the box. If this happens, the 
particles condense, almost motionless, into the lowest quantum state, 
forming a Bose-Einstein condensate. 
So, the Bose-Einstein condensation is a macroscopic quantum phenomenon 
and, thus, similar to the superconducting condensation.  

For many decades physicists dreamt of cooling a sufficiently large number 
of ordinary atoms to low enough temperatures to undergo the Bose-Einstein 
condensation spontaneously. During 1995 this was accomplished by three 
groups acting independently. The first Bose-Einstein condensate was 
formed by using rubidium atoms cooled to 
2 $\times$ 10$^{-9}$ K. 

The superconducting and Bose-Einstein condensates have much in 
common but also a number of differences. Let us start with their 
similarities. Firstly, the superconducting and Bose-Einstein 
condensations are both quantum phenomena occurring on a macroscopic 
scale. Thus, every Bose-Einstein condensate exhibits most of the 
superconducting-state properties. Secondly, the 
superconducting and Bose-Einstein condensations both occur in 
momentum space, not in real space. What is the difference between 
a condensation in momentum space and one in real space? For 
example, the vapor-liquid transition is a condensation in ordinary 
space. After the transition, the average distance between particles (atoms 
or molecules) is changed---becomes smaller when the vapor condenses and 
larger when the liquid evaporates. So, if a condensation takes place in 
real space, there may be some noticeable changes in the system 
(second-order phase transitions occurring in real space, if such exist, 
are not accompanied by changes in real space). On the other 
hand, if a condensation occurs in momentum space there are no changes 
in ordinary space. In the aforementioned example of the Bose-Einstein 
condensation occurring in the box, after the condensation, the mean distance 
between particles remains the same. 

The superconducting and Bose-Einstein 
condensates have two major 
differences. In spite of the fact that the superconducting and Bose-Einstein 
condensates are both quantum states, they, however, have ``different goals to 
achieve.'' Through the Bose-Einstein condensation 
bosons assume to 
reach the lowest energy level existing in the system (see Fig. 2). At the same 
time, the Cooper pairs try to descend below 
the Fermi level 
as deeply as possible, generating an energy gap. The second 
difference is that a Bose-Einstein condensate consists of real bosons, while a 
superconducting condensate comprises composite bosons. To summarize, 
the two condensates---superconducting and Bose-Einstein---have common 
quantum properties, but also, they have a few differences. 

In conventional superconductors, the onset of phase coherence occurs due 
to the overlap of Cooper-pair wavefunctions. In a sense, it is a passive 
process because the overlap of wavefunctions does not generate an 
order parameter---it only makes the Cooper-pair wavefunctions be in 
phase. This means that in order to form a superconducting condensate, 
the Cooper pairs in conventional superconductors must be 
paired in momentum space, not in ordinary space. However, this may not be the 
case for unconventional superconductors where the onset of long-range 
phase coherence occurs due to not the overlap of Cooper-pair 
wavefunctions but due to another ``active'' process. As a consequence, 
if the onset of phase coherence in unconventional superconductors takes 
place in momentum space, it relieves the Cooper pairs of the duty to 
be paired in momentum space. This means that, in unconventional 
superconductors, the Cooper pairs may be formed in real space. 
Of course, they are not required to, but they may. 

If the Cooper pairs in some unconventional superconductors are indeed 
formed in real space, this signifies that the BCS theory and 
the future theory for unconventional superconductors can hardly be unified. 

Let us go back to the second principle of superconductivity. After all these 
explanations, the meaning of this principle should be clear. The transition 
into the superconducting state always occurs in momentum space, and 
this condensation is similar to that predicted by Einstein. 

\section{Third principle of superconductivity} 

If the first two principles of superconductivity, in fact, are just the 
ascertaining of facts and can hardly be used for future predictions, 
the third and fourth principles are better suited for this purpose. 

The third principle of superconductivity is:

\vspace{0.5cm} 
\noindent
Principle 3: {\bf The mechanism of electron 
pairing \\ \hspace*{19mm}and the mechanism of Cooper-pair \\ 
\hspace*{19mm}condensation  must be different} 
\vspace{0.5cm}

The validity of the third principle of superconductivity will be evident 
after the presentation of the fourth principle. Historically, this principle 
was introduced first [2]. 

It is worth to recall that, in conventional superconductors, phonons 
mediate the electron pairing, while the overlap of wavefunctions ensures 
the Cooper-pair condensation. In the unconventional superconductors from 
the third group of superconductors, such as the cuprates, 
organic salts, 
heavy fermions, doped C$_{60}$ etc., phonons also mediate 
the electron pairing, while spin fluctuations are 
responsible for the 
Cooper-pair condensation. So, in all superconductors, the mechanism of 
electron pairing differs from the mechanism of Cooper-pair condensation 
(onset of long-range phase coherence). 
Generally speaking, if in a superconductor, the same ``mediator'' (for 
example, phonons) is responsible for the electron pairing and for the onset of 
long-range 
phase coherence (Cooper-pair condensation), this will simply lead to the 
collapse of superconductivity (see the following section). 

Since in solids, phonons and spin fluctuations have two 
channels---acoustic and optical---{\em theoretically}, it 
is possible that one channel can be responsible for the electron pairing and the 
other for the Cooper-pair condensation. The main problem, however, is that these 
two channels---acoustic and optical---usually compete with one another. So, 
it is very unlikely that such a ``cooperation'' will lead to superconductivity. 

\section{Fourth principle of superconductivity} 

If the first three principles of superconductivity do not deal with 
numbers, the forth principle can be used for making various estimations. 

Generally speaking, a superconductor is characterized by a pairing energy 
gap $\Delta_p$ and a phase-coherence gap $\Delta_c$. 
For genuine (not proximity-induced) superconductivity, the 
phase-coherence gap is proportional to $T_c$: 
\begin{equation} 
2\Delta_c = \Lambda\,k_B T_c, 
\end{equation} 
where $\Lambda$ is the coefficient proportionality [not to be confused 
with the phenomenological parameter $\Lambda$ in the London equations]. 
At the same time, the pairing energy gap is proportional 
to the pairing temperature $T_{pair}$: 
\begin{equation} 
2\Delta_p = \Lambda ' \,k_B T_{pair}. 
\end{equation} 
Since the formation of Cooper pairs must precede the onset of 
long-range phase coherence, then in the general case, $T_{pair} \geq T_c$. 

In conventional superconductors, however, there is only one energy gap 
$\Delta$ which is in fact a pairing gap but proportional to $T_c$:  
\begin{equation} 
2\Delta = \Lambda\,k_B T_c, 
\end{equation} 
This is because, in conventional superconductors, the electron pairing and 
the onset of long-range phase coherence take place at the same 
temperature---at $T_c$. In all known cases, the coefficients $\Lambda$ 
and $\Lambda '$ lie in the interval between 3.2 and 6 (in one heavy 
fermion, $\sim$ 9). Thus, we are now in position to discuss 
the fourth principle of superconductivity:

\vspace{0.5cm} 
\noindent
Principle 4: {\bf For genuine, homogeneous 
supercon- \\ \hspace*{19mm}ductivity, $\Delta_p > \Delta_c > \frac{3}{4}k_BT_c$ 
always \hspace*{19mm}(in conventional superconductors, 
\\\hspace*{19mm}$\Delta > \frac{3}{4}k_BT_c$)
} 
\vspace{0.5cm}

Let us start with the case of conventional superconductors. The reason 
why superconductivity occurs exclusively at low temperatures is the 
presence of substantial thermal fluctuations at high temperatures. 
The thermal energy is $\frac{3}{2}k_BT$. In conventional 
superconductors, the energy of electron binding, $2\Delta$, must be 
larger than the thermal energy; otherwise, the pairs will be broken up 
by thermal fluctuations. So, the energy $2\Delta$ must exceed the 
energy $\frac{3}{2}k_BT_c$. In the framework of the BCS theory, 
the ratio between these two energies, 
$2\Delta /(k_B T_c) \simeq$ 3.52, is well above 1.5. 

In the case of unconventional superconductors, the same reasoning is also 
applicable for the phase-coherence energy gap: 
$2\Delta_c > \frac{3}{2}k_BT_c$. 

We now discuss the last inequality, namely, $\Delta_p > \Delta_c$. In 
unconventional superconductors, the Cooper pairs condense at 
$T_c$ due to their interaction with some bosonic excitations present in the system, 
for example, spin fluctuations. These bosonic excitations 
are directly coupled 
to the Cooper pairs, and the strength of this coupling with each Cooper pair 
is measured by the energy $2\Delta_c$. If the strength of this coupling 
will exceed the pairing energy $2\Delta_p$, the Cooper pairs will 
immediately be broken up. Therefore, the inequality $\Delta_p > \Delta_c$ 
must be valid. 

What will happen with a superconductor if, at some temperature, 
$\Delta_p = \Delta_c$? Such a situation can take place either at $T_c$, 
defined {\em formally} by Eq. (1), or below $T_c$, i.e. inside the 
superconducting state. In both cases, the temperature at which such a 
situation occurs is a critical point, $T_{cp}$. If the temperature 
remains constant, locally there will be superconducting fluctuations due 
to thermal fluctuations, thus, a kind of inhomogeneous superconductivity. If 
the temperature falls, two outcomes are possible (as it usually takes 
place at a critical point). In the first scenario, superconductivity will 
never appear if $T_{cp} = T_c$, or will disappear at $T_{cp}$ if 
$T_{cp} < T_c$. In the second possible outcome, homogeneous 
superconductivity may appear. The final result depends completely on 
bosonic excitations that mediate the electron pairing and that responsible 
for the onset of phase coherence. The interactions of these excitations 
with electrons and Cooper pairs, respectively, vary with temperature. If, 
somewhat below $T_{cp}$, the strength of the pairing binding increases 
{\em or/and} the strength of the phase-coherence adherence decreases, 
homogeneous superconductivity will appear. In the opposite 
case, superconductivity will never appear, or disappear at $T_{cp}$. 
It is worth noting that, in principle, superconductivity may reappear at 
$T < T_{cp}$.

The cases of disappearance of superconductivity below $T_c$ are well 
known. However, it is assumed that the cause of such a disappearance is 
the emergence of a ferromagnetic order. As well known,
the Chevrel phase HoMo$_6$S$_8$ is superconducting only between 2 and 
0.65 K. The erbium rhodium boride ErRh$_4$B$_4$ superconducts only 
between 8.7 and 0.8 K. The cuprate \linebreak Bi2212 doped by Fe atoms was seen 
superconducting only between 32 and 31.5 K [4]. The so-called 
$\frac{1}{8}$ anomaly in the cuprate LSCO, discussed in Chapter 3, 
is caused apparently by {\em static} magnetic order [2] which may result 
in the appearance of a critical point where $\Delta_p \simeq \Delta_c$. 

It is necessary to mention that the case $\Delta_p = \Delta_c$ must not 
be confused with the case $T_{pair} = T_c$. There are unconventional 
superconductors in which the electron pairing and the onset of phase 
coherence occur at the same temperature, i.e. $T_{pair} \simeq T_c$. 
This, however, does not mean that $\Delta_p = \Delta_c$ because 
$\Lambda \neq \Lambda '$ in Eqs. (1) and (2). Usually, 
$\Lambda ' > \Lambda$. For example, in hole-doped cuprates, 
$2\Delta_p/k_BT_{pair} \simeq$ 6 and, depending on the cuprate, 
$2\Delta_c/k_BT_c =$ 5.2--5.9. 

Finally, let us go back to the third principle of superconductivity to show 
its validity. The case in which the same bosonic excitations mediate the 
electron pairing {\bf and} the phase coherence is equivalent to the case 
$\Delta_p = \Delta_c$ discussed above. Since, in this particular case, the 
equality $\Delta_p = \Delta_c$ is independent of temperature, the 
occurrence of homogeneous superconductivity is impossible. 

\section{Proximity-induced superconductivity} 

The principles considered above are derived 
for genuine superconductivity. 
By using the same reasoning as that in the previous section for 
proximity-induced superconductivity, one can obtain a useful result, 
namely, that $2\Delta \sim \frac{3}{2}k_B T_c$, meaning that the energy 
gap of proximity-induced superconductivity should be somewhat larger 
than the thermal energy. Of course, to observe this gap for example in 
tunneling measurements may be not possible if the density of induced 
pairs is low. This case is reminiscent of gapless superconductivity 
[5]. Hence, we may argue 
that 

\vspace{0.5cm} 
\noindent 
\hspace*{9mm} {\bf For proximity-induced superconductivity, 
\\ \hspace*{10mm}at low temperature, $2\Delta_p \geq \frac{3}{2} k_B T_c$ 
} 
\vspace{0.6cm}

One should however realize that this is a general statement; the final 
result depends also upon the material and, in the case of thin films, on 
the thickness of the normal layer. 

What is the maximum critical temperature of BCS-type superconductivity? 
In conventional superconductors, $\Lambda =$ 3.2--4.2 in Eq. (3). Among 
conventional superconductors, Nb has the maximum energy gap, 
$\Delta \simeq$ \linebreak 1.5 meV. Then, taking $\Delta_{max}^{BCS} \approx$ 2 meV 
and using $\Lambda$ = 3.2, we have $T_{c,max}^{BCS} = 
2\Delta_{max}^{BCS}/3.2k_B \approx$ 15 K for conventional 
superconductors. Let us now estimate the maximum critical temperature 
for induced superconductivity of the BCS type in a material with a 
strong 
electron-phonon interaction. In such materials, genuine superconductivity 
(if exists) is in the strong coupling regime and 
characterized by $\Lambda \simeq$ 4.2 
in Eq. (3). Assuming that the same strong coupling regime is also applied 
to the induced superconductivity with $2\Delta \sim 1.5k_B T_c^{ind}$ and 
that, in the superconductor which induces the Cooper pairs, 
$\Delta_p \gg$ 2 meV, one can then obtain that 
$T_{c,max}^{ind} \sim$ 15 K $\times\,\frac{4.2}{1.5} \simeq$ 42 K. 

If the superconductor which induces the Cooper pairs is of the BCS type, 
the value $\Delta_{max}^{ind}$ = 2 meV can be used to estimate 
$T_{c,max}^{ind}$ independently. Substituting the value of 2 meV into 
$2\Delta \sim 1.5k_B T_c^{ind}$, we have $T_{c,max}^{ind} \simeq$ 31 K 
which is lower than 42 K. 

In second-group superconductors which are characterized by the presence 
of two superconducting subsystems, the critical temperature never 
exceeds 42 K. For example, in MgB$_2$, $T_c$ = 39 K and, for the smaller 
energy gap, $2\Delta_s \simeq 1.7k_BT_c$ [2]. At the same 
time, for the larger energy gap in MgB$_2$, 
$2\Delta_L \simeq 4.5k_BT_c$ or $\Delta_L \simeq$ 7.5 meV. Then, on 
the basis of the estimation for $T_{c,max}^{ind}$, it is more or less 
obvious that, in MgB$_2$, one subsystem with genuine superconductivity 
(which is low-dimensional), having $\Delta_L \simeq$ 7.5 meV, induces 
superconductivity into another subsystem and the latter one controls the 
bulk $T_c$. 

The charge carriers in compounds of the first and second groups of 
superconductors are electrons. Is there hole-induced superconductivity? 
Yes. At least one case of hole-induced superconductivity is known: in the 
cuprate YBCO, the CuO chains become superconducting 
due to the proximity effect. The value of the superconducting energy gap 
on the chains in YBCO is well documented; in optimally doped YBCO, it is 
about 6 meV [2]. Using $T_{c,max} =$ 93 K for YBCO and $\Delta \sim$ 6 meV, 
one obtains $2\Delta/k_BT_c \simeq$ 1.5. This result may indicate that 
the bulk $T_c$ in YBCO is controlled by induced superconductivity on the 
CuO chains.

\end{document}